\documentclass[12pt]{article}
\usepackage{amssymb}
\usepackage{graphicx}
\usepackage{indentfirst}
\usepackage{amsmath}
\usepackage{braket}

\begin{document}

\begin{titlepage}
\begin{center}

\vspace*{25mm}

{\Large\bf Extraction of $|V_{cb}|$ from two-body hadronic B decays}
\vspace*{25mm}

{\large
Noriaki Kitazawa\footnote{noriaki.kitazawa@tmu.ac.jp},
Kyo-suke Masukawa\footnote{masukawa-kyosuke@ed.tmu.ac.jp} and
Yuki Sakai\footnote{sakai-yuki1@ed.tmu.ac.jp}
}
\vspace{10mm}

Department of Physics, Tokyo Metropolitan University,\\
Hachioji, Tokyo 192-0397, Japan\\

\vspace*{25mm}

\begin{abstract}

We propose a method of extracting the Cabibbo-Kobayashi-Maskawa matrix element $|V_{cb}|$
 from two-body hadronic decay processes of $B\to DK$
 with precisely determined form factors of $B$ meson semi-leptonic decays.
The amplitude $\mathcal{M}(\bar{B}^0 \to D^+ K^-)$
 which does not include the effect of hadronic final state interactions
 can be theoretically evaluated by using factorization
 and form factors of semi-leptonic B decays.
We can obtain all the amplitudes
 in an isospin relation
 $\mathcal{A}(B^-\to D^0K^-)
 =\mathcal{A}(\bar{B}^0\to D^+K^{-})+\mathcal{A}(\bar{B}^0\to D^{0}\bar{K}^0)$
 including the effect of hadronic final state interactions as well as $|V_{cb}|$
 using the experimental data of branching fractions of these three processes
 with a truncation of the states which contribute to the hadronic final state interactions.
The extracted value of $|V_{cb}|$ is $(37\pm 6)\times 10^{-3}$.
The decay processes of $B\to DK^{*}$ and $B\to D^{*}K$ can also be used in the same way
 and the extracted values of $|V_{cb}|$ are $(41\pm 7)\times 10^{-3}$
 and $(42\pm 9)\times 10^{-3}$, respectively.
This method becomes possible
 by virtue of recent precise determinations of the form factors of semi-leptonic B decays.
The uncertainties of $|V_{cb}|$ by this method are expected to be reduced
 by the results of future B-factory experiments and lattice calculations.

\end{abstract}

\end{center}
\end{titlepage}

\setcounter{footnote}{0}

\section{Introduction}
\label{introduction}

The precise determination of Cabibbo-Kobayashi-Maskawa (CKM) matrix elements
 \cite{Cabibbo:1963yz,Kobayashi:1973fv}
 is one of the approaches to test the Standard Model
 and to search the physics beyond the Standard Model. 
The Standard Model predicts the unitarity relation of CKM matrix
\begin{equation}
 \sum_{i=u,c,t} V^{*}_{ib}V_{id} = 0
\label{unitarity relation}
\end{equation}
 which gives a triangle in a complex plane. 
The existence of the physics beyond the Standard Model may violate this relation.
The sides and angles of this triangle will be precisely measured
 using various decay processes of B mesons in future B-factories
 \cite{Abe:2010gxa,Bediaga:2012py}. 
In this work we focus on the determination of $|V_{cb}|$.

There are mainly two methods to extract the value of $|V_{cb}|$ from semi-leptonic B meson.
The method using inclusive decay data gives 
 $|V_{cb}|=(42.00\pm0.65)\times10^{-3}$ \cite{Gambino:2016jkc},
 and the method using exclusive decay data gives
 $|V_{cb}|=(38.71\pm0.75)\times10^{-3}$ \cite{Amhis:2016xyh}.
Though the difference of these values are within $3.3 \sigma$, it is a problem
 in understanding non-perturbative physics of QCD.\footnote{
It has been pointed out that
 this problem can not be solved by New Physics \cite{Crivellin:2014zpa}.}
In fact, it has been pointed out that
 the proper parameterization of form factors is important \cite{Bigi:2016mdz,Bigi:2017njr}. 
In order to analyze exclusive decay processes $B\to D^{(*)}l\nu$
 with a small amount of data,
 the Caprini-Lellouch-Neubert (CLN) parameterization of form factors \cite{Caprini:1997mu}
 is precise enough.
However,
 with much more data recently provided by Belle collaboration,
 not only $q^2$-distributions but also angular-distributions \cite{Abdesselam:2017kjf},
 the Boyd-Grinstein-Lebed (BGL) parameterization of form factor \cite{Boyd:1997kz}
 is better than the CLN parametrization,
 because the CLN parametrization may include about 10\% errors
 from the absence of $\mathcal{O}(1/m_{c,b}^2)$ corrections \cite{Bigi:2017njr}.
Since the accuracy of recent lattice QCD results
 \cite{Lattice:2015rga,Na:2015kha} is typically of the order of 1\%, 
 we need to use theoretical frameworks with correspondingly high precisions.\footnote{
The error with CNL parameterization
 comes from an excessive reduction of the number of parameters in form factors
 by using heavy quark symmetry.
In fact improvements are possible by including higher order corrections
 (for example, see \cite{Bernlochner:2017jka}).}

In this work we intend to provide another method to extract $|V_{cb}|$,
 which may give new information to the above conflict in future.
We propose that
 the hadronic decays of B mesons,
 especially two-body decays of $B\to DK$, $B\to DK^{\ast}$ and $B\to D^{\ast}K$,
 can be used to extract precise value of $|V_{cb}|$ in future.  
The amplitude $\mathcal{M}(\bar{B}^0\to D^+K^-)$
 which does not include the effects of hadronic final state interactions
 can be theoretically evaluated by using the factorization,
 the form factors of semi-leptonic B decays and decay constant of $K^-$ meson.
The form factors of semi-leptonic B decays are precisely determined
 by the latest Belle data \cite{Abdesselam:2017kjf,Glattauer:2015teq}
 and the latest lattice QCD results \cite{Lattice:2015rga,Na:2015kha}
 with the BGL parameterizations in \cite{Bigi:2016mdz}.
The isospin symmetry provides a relation
\begin{equation}
 \mathcal{A}(B^-\to D^0K^-)
 = \mathcal{A}(\bar{B}^0\to D^+K^{-})+\mathcal{A}(\bar{B}^0\to D^{0}\bar{K}^0)
\label{isospin relation}
\end{equation}
 including the effects of hadronic final state interactions.
We can extract these three amplitudes as well as the value of $|V_{cb}|$ by using
 the amplitude $\mathcal{M}(\bar{B}^0\to D^+K^-)$
 and the experimental values of three branching fractions
 $\mathcal{B}(B^-\to D^0K^-)$, $\mathcal{B}(\bar{B}^0\to D^+K^{-})$ and
 $\mathcal{B}(\bar{B}^0\to D^{0}\bar{K}^0)$ in \cite{Patrignani:2016xqp}.
In this procedure
 we need to truncate the states which contribute final state interactions:
 not including all the possible states, but including only two-body $DK$ states.
The processes of $B\to DK^{*}$ and $B\to D^{*}K$ can also be used in the same way.
For $B\to D^{*}K$,
 we use the form factor obtained by the CLN parameterization
 with latest data by Belle collaboration \cite{Abdesselam:2017kjf}.

We emphasize that this method becomes possible
 only with recent precise determination of all the form factors of semi-leptonic B decay.
More precise experimental data of the branching fractions of two-body hadronic B decays
 give more precise value of $|V_{cb}|$.
This method can be understood as an intermediate approach
 between inclusive and exclusive determination of $|V_{cb}|$,
 since it requires to use several exclusive B-decay modes.
It may be possible that
 this method will play an important role in the problem of $|V_{cb}|$ determinations
 with the results of future B-factory experiments and future precise lattice calculations,
 if the validity of the truncation of the states in final state interactions
 is established.
In other words,
 once the value of $| V_{cb} |$ is precisely determined
 with semi-leptonic decays without any conflicts,
 this method will provide useful information
 to understand the final state interactions in two-body hadronic B decays.

In the next section 
 we investigate the amplitudes of $B\to DK$ processes in detail,
 and propose a procedure to extract the value of $|V_{cb}|$.
We also show that
 the same procedure applies to the processes of $B\to DK^{\ast}$ and $B\to D^{\ast}K$.
In section \ref{numerical}
 we provide the numerical analyses of extracting the value of $|V_{cb}|$
 from two-body hadronic B decays by our procedure.
In section \ref{conclusions} we provide a summary and discussion.


\section{Two-body hadronic decays of B mesons}
\label{model}

\begin{figure}[tbp]
\centering
\includegraphics[width=110mm]{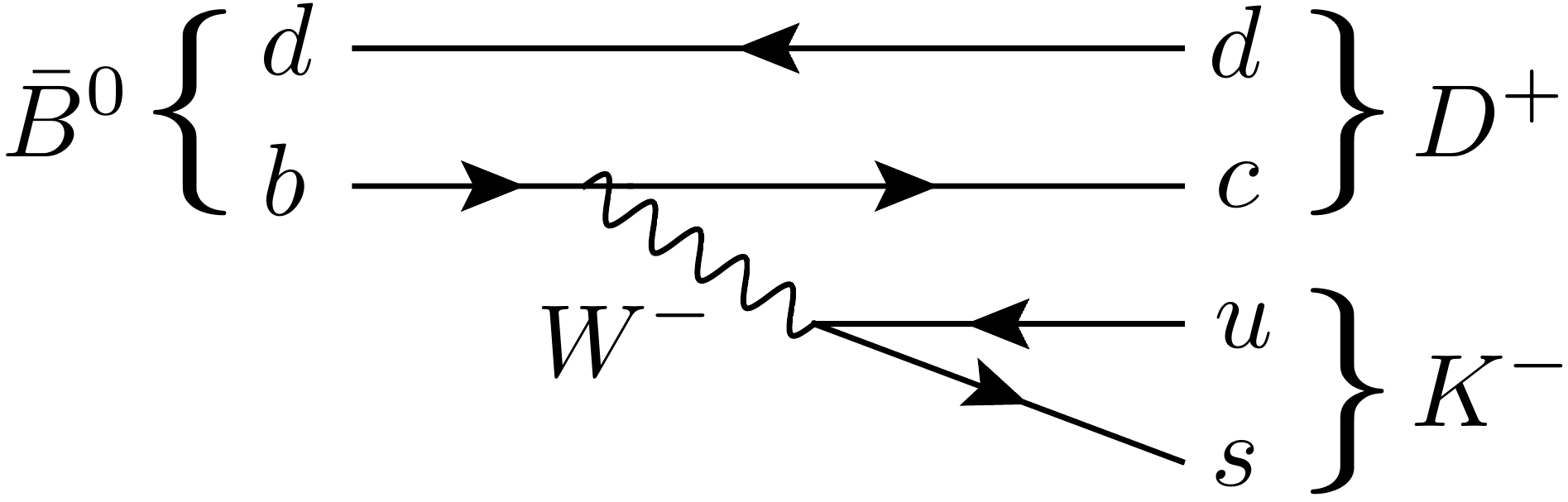}
\caption{
Tree amplitude.
For example, $M_1$ and $M_2$ meson are $D^+$ and $K^-$, respectively.
}
\label{fig:TreeDiagram}
\end{figure}

\begin{figure}[tbp]
\centering
\includegraphics[width=110mm]{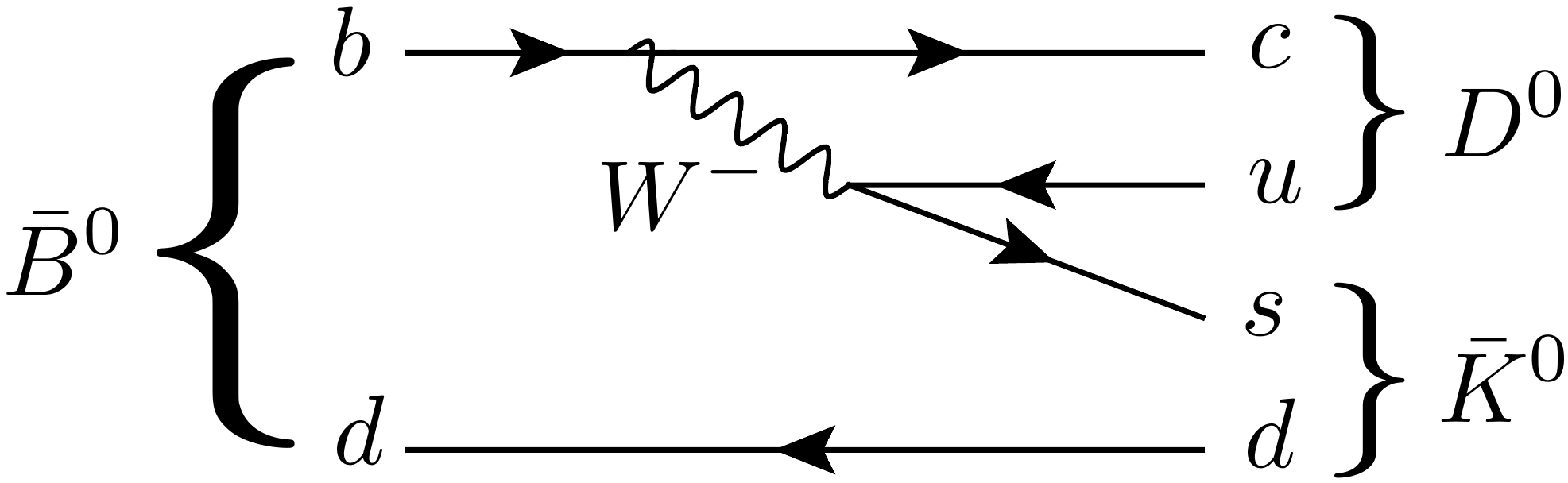}
\caption{
Color-suppressed amplitude.
For example, $M_1$ and $M_2$ meson are $D^0$ and $\bar{K}^0$, respectively.
}
\label{fig:ColorDiagram}
\end{figure}

\begin{figure}[tbp]
\centering
\includegraphics[width=110mm]{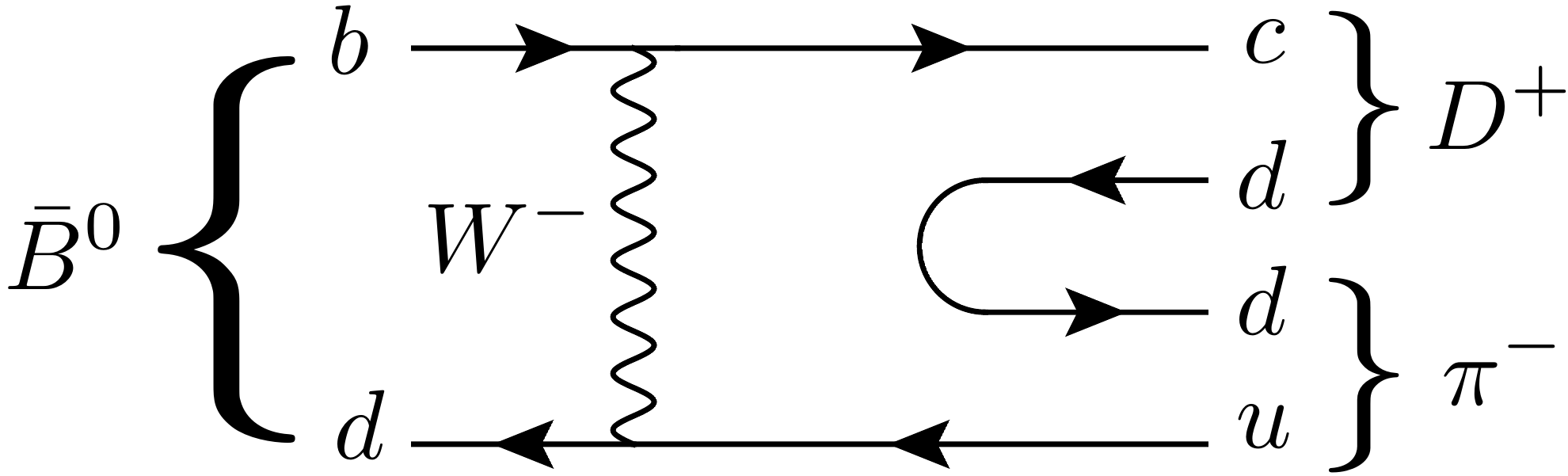}
\caption{
Exchange amplitude where a quark-antiquark pair creation occurs.
For example, $M_1$ and $M_2$ meson are $D^+$ and $\pi^-$, respectively.
}
\label{fig:ExchangeDiagram}
\end{figure}

\begin{figure}[tbp]
\centering
\includegraphics[width=110mm]{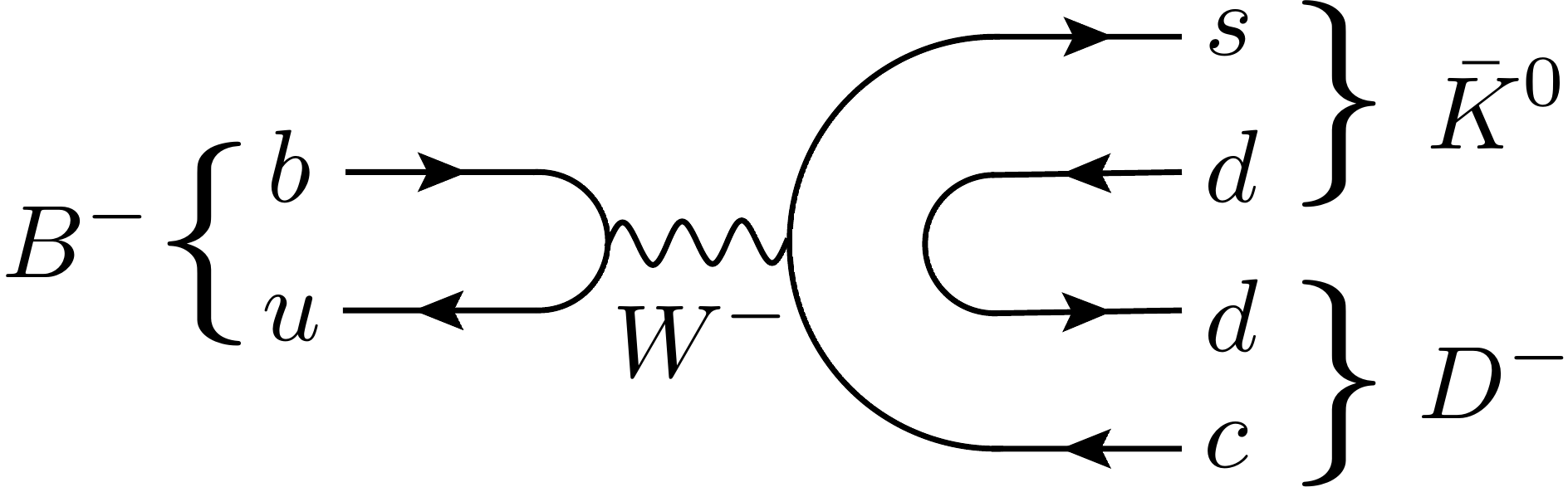}
\caption{
W-annihilation amplitude.
For example, $\bar{B}$, $M_1$ and $M_2$ meson are $B^-$, $D^-$ and $\bar{K}^0$, respectively.
}
\label{fig:AnnihilationDiagram}
\end{figure}

Consider the hadronic two-body decay processes $\bar{B} \to M_1 M_2$,
 where $M_1$ and $M_2$ indicate $D$ mesons and $K$ or $\pi$ mesons, respectively.
The quark-level Feynman diagrams of these decays
 are classified into four topological types \cite{Chiang:2007bd,Zhou:2015jba}.
The amplitudes from the diagrams corresponding to each topological type are called as follows.
(1) Tree amplitudes $T$:
 the diagrams have $b \to c$ weak current
 with the light degrees of freedom as spectator antiquarks of $\bar{B}$ and $M_1$ mesons,
 and the $W$ boson decays into the light quark-antiquark pairs which constitute $M_2$ meson
 (see Fig.\ref{fig:TreeDiagram}).
(2) Color-suppressed amplitudes $C$: 
 the $W$ boson decays into the light quark-antiquark pairs, 
 and the antiquark is included in $M_1$ meson as the spectator of $c$ quark,
 and the quark constitutes $M_2$ meson with the light degrees of freedom
 in $\bar{B}$ meson (see Fig.\ref{fig:ColorDiagram}).
(3) Exchange amplitudes $E$:
 the exchange of the $W$ boson changes the flavor of spectator of $\bar{B}$,
 and light quark-antiquark pair creation from gluons completes two mesons
 (see Fig.\ref{fig:ExchangeDiagram}).
(4) W-annihilation amplitudes $A$: 
 the $\bar{B}$ meson decays to a $W$ boson
 and the $W$ boson decays into a charm antiquark and a light quark, 
 and they become constituents of $M_{1}$ and $M_{2}$
 with a light quark and and light antiquark from gluons, respectively
 (see Fig.\ref{fig:AnnihilationDiagram}).
In this paper we do not consider the process which contains the contribution of $A$,
 since it does not include $|V_{cb}|$ and it is rather relevant to $|V_{ub}|$.

\begin{figure}[tbp]
\centering
\includegraphics[width=100mm]{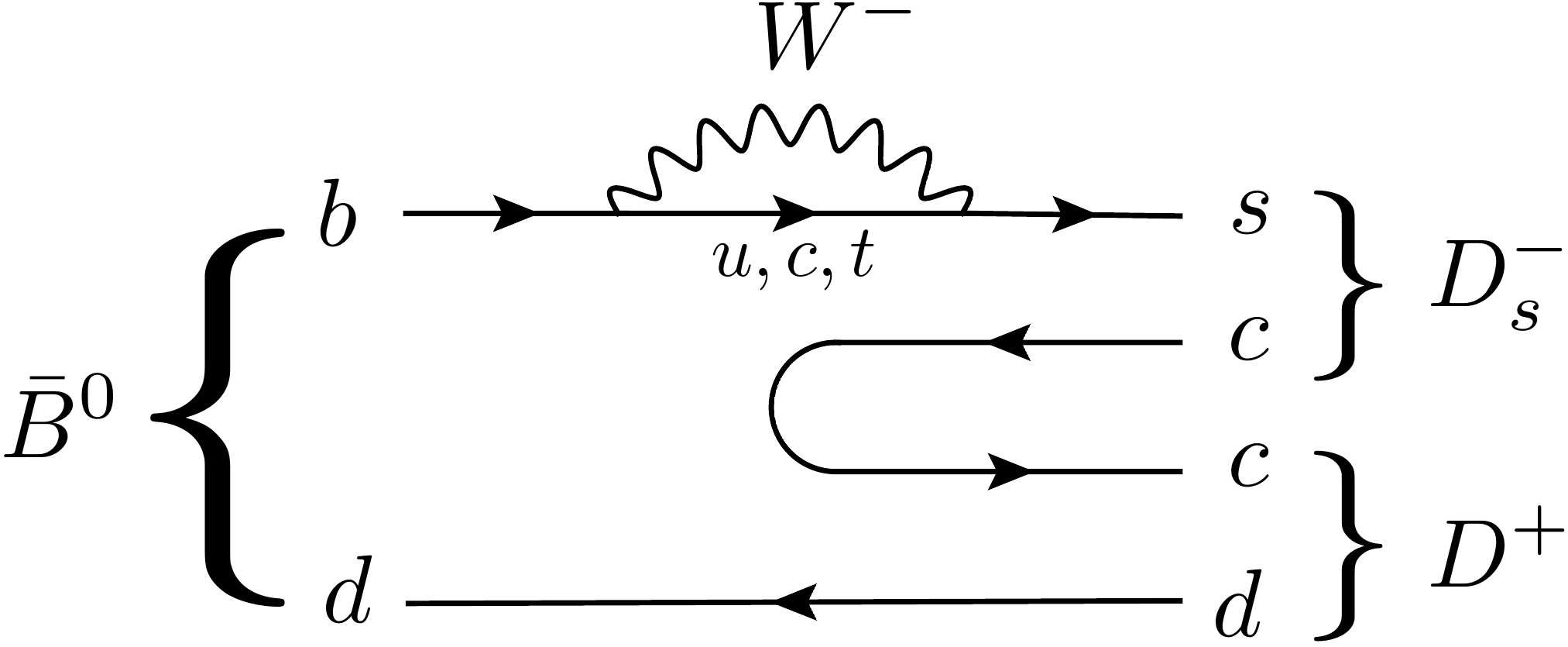}
\caption{
Penguin diagram.
The process of $\bar{B}^0\to D_s^- D^+$ is shown as an example.
}
\label{fig:Penguin}
\end{figure}

In table \ref{table:PPchannel} we summarize
 all the hadronic two-body decays of $\bar{B}^0$ and $B^-$ mesons,
 which include $b \to c$ transition,
 and the topologies of corresponding amplitudes. 
\begin{table}[tbp]
\centering
 \begin{tabular}{rccccc} \hline \hline
 \multicolumn{1}{r}{Decay mode} & \multicolumn{3}{c}{topologies} &
 \multicolumn{1}{c}{Penguin} &
 \multicolumn{1}{c}{Fraction $(\Gamma_i/\Gamma)$}\cite{Patrignani:2016xqp}
 \\
 \hline
    $\bar{B}^0\to D^+\pi^-$ & T & & E & & $(2.52\pm 0.13)\times10^{-3}$ \\
    $D^0\pi^0$ &  & C & E & & $(2.63\pm 0.14)\times 10^{-4}$ \\
    $D_s^{+}K^-$ &  & C & E & & $(2.7\pm 0.5)\times 10^{-5}$ \\
    $D^{+}K^-$ & T & &  &  & $(1.86\pm 0.20)\times10^{-4}$ \\
    $D^0\bar{K}^0$ &  & C &  &  & $(5.2\pm 0.7)\times10^{-5}$ \\
    $D^{+}D^{-}$ & T & & E & yes & $(2.11\pm 0.18)\times10^{-4}$ \\ 
    $D^+D_s^{-}$ & T & &  & yes & $(7.2\pm 0.8)\times10^{-3}$ \\ 
    ${B^-}\to D^0\pi^-$ & T & C &  &  & $(4.80\pm 0.15)\times10^{-3}$ \\
    $D^0K^-$ & T & C &  &  & $(3.74\pm 0.16)\times10^{-4}$ \\
    $D^0D^-$ & T & C & & yes & $(3.8\pm 0.4)\times10^{-4}$ \\
    $D^0D_s^-$ & T & C & & yes & $(9.0\pm 0.9)\times10^{-3}$ \\
 \hline \hline
 \end{tabular}
\caption{
Two-body hadronic decays and their amplitudes.
Note that the decay mode $\bar{B}^0\to D^+K^-$ is the only mode
 which is described by the diagram of Tree topology only.
The contributions of penguin diagrams are also listed. 
}
\label{table:PPchannel}
\end{table}
In general
 several diagrams with different topologies contribute to the amplitudes for each decay process.
We see that
 the amplitude of $\bar{B}^0\to D^+K^-$ consists of a single diagram of topology $T$.  
The penguin diagrams (see Fig.\ref{fig:Penguin}) contribute only to the amplitude of $B \to DD$.
For example,
 the amplitude of $\bar{B}^0\to D^+D_s^-$ consists of $T$ with a pollution by a penguin diagram.
Until the size of the contribution of the penguin diagram is clarified,
 we can not use $\bar{B}^0\to D^+D_s^-$ to extract $|V_{cb}|$ precisely. 

We focus on the two-body decay $\bar{B}^0\to D^+K^-$
 which is described only by the diagram of topology $T$.
The effective weak Hamiltonian \cite{Buchalla:1995vs} for the decay is
\begin{equation}
H_{\text{eff}} = \eta_{\text{EW}}\frac{G_\text{F}}{\sqrt{2}} V_{cb}V^*_{us} \left[ C_1(\mu)O_1(\mu) + C_2(\mu)O_2(\mu)   \right] + \text{h.c.},
\end{equation}
 where $\eta_{\text{EW}}$ is the electroweak correction
 which represents the effects of short-distance QED correction.
The factors $C_{1,2}$
 are the Wilson coefficients and $O_{1,2}$ are the current-current operators:
\begin{eqnarray}
 O_1 &=& \bar{c}^{\alpha}\gamma_{\mu}(1-\gamma_5)b_{\beta}
         \bar{s}^{\beta}\gamma^\mu(1-\gamma_5)u_{\alpha},  \\
 O_2 &=& \bar{c}^{\alpha}\gamma_{\mu}(1-\gamma_5)b_{\alpha}
         \bar{s}^{\beta}\gamma^\mu(1-\gamma_5)u_{\beta},
\end{eqnarray}
 where $\alpha$ and $\beta$ are color indices.
The amplitude is given by the matrix element 
\begin{equation}
 \mathcal{A}(\bar{B}^0\to D^+K^-)
  = \eta_{\text{EW}} \frac{G_\text{F}}{\sqrt{2}} V_{cb}V^*_{us} a_1(\mu)
    \bra{D^{+}(p')K^-(p_K)}
     [\bar{c} \gamma_{\mu}(1-\gamma_5)b][\bar{s}\gamma^{\mu}(1-\gamma_5)u]
    \ket{\bar{B}^0(p)},
\label{matrix-element-BDK}
\end{equation}
 where $p, p'$ and $p_K$ are 4-momenta of $\bar{B}^0,D^+$ and $K^-$, respectively.
The momentum $p_{K}$ satisfies $q^2=(p-p')^2=p_K^2$.
The factor $a_1(\mu)=C_2(\mu)+C_1(\mu)/3$
 represents the effects of short-distance QCD correction
 including short-distance non-factorizable QCD effects.
The amplitude $\mathcal{A}(\bar{B}^0\to D^+K^-)$ includes also the effects of
 non-factorizable hadronic final state interactions (or rescattering effects)
 which are non-perturbative QCD effects.
Now, we introduce the amplitude $\mathcal{M}(\bar{B}^0\to D^+K^-)$
 which does not include the effects of hadronic final state interactions.
The amplitude is given by
 factorizing the matrix element in eq.(\ref{matrix-element-BDK}),
 because only the diagram of topology $T$ contributes.
The final state is written by two independent asymptotic states of $D^+$ and $K^-$ mesons,
 because we have temporarily neglected the effects of final state interactions,
 or long-distance non-factorizable QCD effects.
The amplitude is written as
\begin{eqnarray}
 \mathcal{M}(\bar{B}^0\to D^+K^-)
  &=& \eta_{\text{EW}} \frac{G_\text{F}}{\sqrt{2}} V_{cb}V^*_{us} a_1(\mu)
      \bra{D^{+}(p')} \bar{c}\gamma_{\mu}(1-\gamma_5)b \ket{\bar{B}^0(p)}
\nonumber \\
  && \quad \times \bra{K^-(p_K)} \bar{s}\gamma^\mu(1-\gamma_5)u \ket{0}.
\end{eqnarray}
The $B \to D$ part of the matrix element is given by 
\begin{equation}
 \bra{D(p')} \bar{c} \gamma_{\mu}(1-\gamma_5)b \ket{\bar{B}(p)}
  = f_{+}(q^2)(p+p')_{\mu} + f_{-}(q^2)(p-p')_{\mu},
\label{eq:FormFactors}
\end{equation}
 where $f_{\pm}(q^2)$ are form factors of semi-leptonic decay of $\bar{B}^0$. 
Another matrix element is described as
\begin{equation}
 \bra{0}\bar{u}\gamma_{\mu}(1-\gamma_{5})s \ket{K(p_K)} = i p_{K\mu} f_{K^{\pm}},
\end{equation}
 where $f_{K^{\pm}}= 155.6\pm0.4$MeV \cite{Patrignani:2016xqp}
 is the decay constant of $K^{\pm}$ mesons.
The absolute value of the amplitude is written as
\begin{equation}
|\mathcal{M}(\bar{B}^0\to D^+K^-)|
 = \eta_{\text{EW}}\frac{G_{\text{F}}}{\sqrt{2}} |V_{cb}V_{us}^*| a_1(\mu)
   (m_{\bar{B}^0}^2-m_{D^+}^2) f_{K^{\pm}} f_{0}(m_{K^{\pm}}^2), 
\label{eq:amplitudeBDK}
\end{equation}
 where 
\begin{equation}
 f_{0}(q^2) = f_{+}(q^2) + \frac{q^2}{m_B^2-m_D^2}f_{-}(q^2)
\end{equation}
 and the function of $f_{0}(q^2)$
 is precisely determined in \cite{Bigi:2016mdz} for all possible $q^2$ region.
Notice that this amplitude depends only on one form factor $f_{0}(q^2)$.
If we could neglect the effect of hadronic final state interactions,
 the value of $|V_{cb}|$ could be straightforwardly extracted
 from the data of the decay rate in \cite{Patrignani:2016xqp},
 since the decay rate is simply described as
\begin{equation}
 \Gamma(\bar{B}^0\to D^+K^-)\vert_{\rm no \, FSI}
  = \frac{p^*}{8\pi m_{B}^2} |\mathcal{M}(\bar{B}^0\to D^+K^-)|^2,
\end{equation}
 where 
\begin{equation}
 p^* = \frac{1}{2m_{\bar{B}^0}}
  \sqrt{\{ m_{\bar{B}^0}^2-(m_{D^+}+m_{K^-})^2\}
        \{m_{\bar{B}^0}^2-(m_{D^+}-m_{K^-})^2\}}.
\end{equation}
We obtain the value of $|V_{cb}| = (32.0 \pm 1.9) \times 10^{-3}$ 
 which is inconsistent with the values determined by the inclusive and exclusive methods
 with semi-leptonic decays. 
This result indicates the failure of ``naive factorization'' and shows that
 the effect of hadronic final state interactions can not be ignored
 and it is important to extract $|V_{cb}|$ from hadronic two-body B decays.\footnote{
We have also neglected
 the effect of non-factorizable spectator quark scattering
 which violate the factorization \cite{Beneke:2003zv}.
It has been shown in \cite{Beneke:2000ry} that
 the effect is small in heavy quark mass limit in case that
 the spectator quark goes to a heavy meson, like in $\bar{B}^0\to D^+K^-$ decay.
We neglect the effect in this work
 keeping in mind that we will need to include the small effect
 with precise experimental data in future.
}

\begin{figure}[tbp]
\centering
\includegraphics[width=120mm]{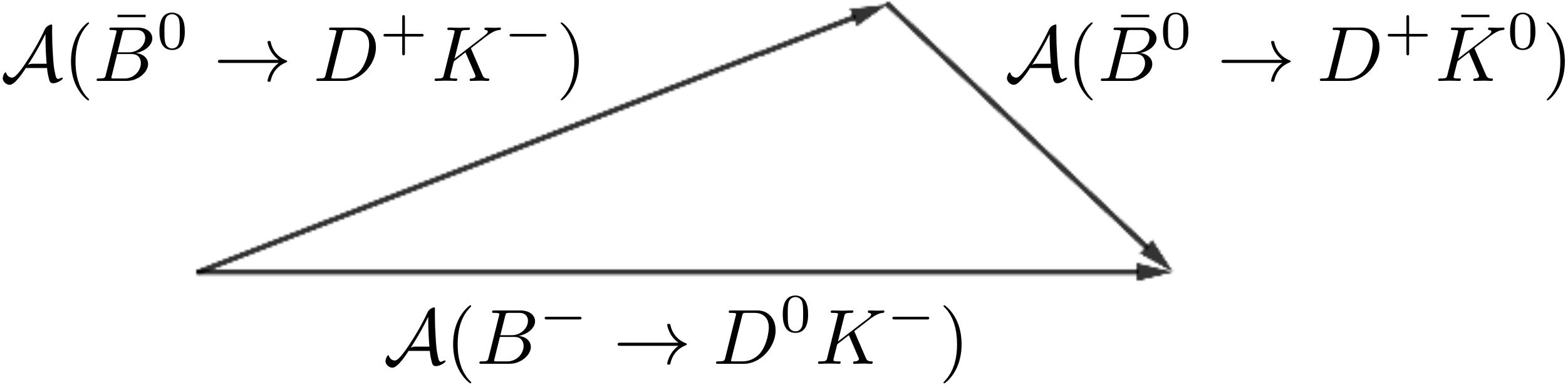}
\caption{
Isospin relation of three amplitudes in a complex plane.
We choose the direction of the amplitude $\mathcal{A}(B^-\to D^0K^-)$
 as that of the the real axis. 
Non-zero strong phases of  $\delta_{0}$ and $\delta_{1}$ mean non-zero area of this triangle.
}
\label{fig:TriangleRelation}
\end{figure}

In order to consider the effect of hadronic final state interactions, 
 we introduce a relation between decay amplitudes which follows from isospin symmetry.
The amplitudes of $B^-\to D^0K^-$, $\bar{B}^0\to D^+K^-$ and $\bar{B}^0\to D^0 \bar{K}^0$ are related by isospin symmetry as
\begin{equation}
\mathcal{A}(B^-\to D^0K^-)
 = \mathcal{A}(\bar{B}^0\to D^+K^-) + \mathcal{A}(\bar{B}^0\to D^0 \bar{K}^0).
\end{equation} 
We expect that this relation should be satisfied within 1\% accuracy,
 because the isospin breaking effect should be proportional to
 $(m_d - m_u)/\Lambda_{\rm QCD} \sim 0.02$ or $\alpha/\pi \sim 0.002$.
We can represent this relation as a triangle on a complex plane
 (see Fig.\ref{fig:TriangleRelation}).
The isospin decomposition of these amplitudes are given by 
\begin{eqnarray} 
 \mathcal{A}(B^-\to D^0K^-)
  &=& A_1 = |A_{1}|e^{i\delta_{1}},
\label{eq:amplitudes-1}
\\
 \mathcal{A}(\bar{B}^0\to D^+K^-)
  &=& \frac{1}{2} (A_1+A_0)  = \frac{1}{2}(|A_1|e^{i\delta_{1}}+|A_0|e^{i\delta_{0}})
  \equiv \frac{1}{2}(|A_1|+|A_0|e^{i\delta_{s}}) e^{i\delta_{1}},
\label{eq:amplitudes-2}
\\
 \mathcal{A}(\bar{B}^0\to D^0\bar{K}^0)
  &=& \frac{1}{2}(A_1-A_0) = \frac{1}{2}(|A_1|e^{i\delta_{1}}-|A_0|e^{i\delta_{0}})
  = \frac{1}{2}(|A_1|-|A_0|e^{i\delta_{s}}) e^{i\delta_{1}},
\label{eq:amplitudes-3}
\end{eqnarray}
 where $\delta_{0}$ and $\delta_{1}$
 are phases by the effect of hadronic final state interactions
 of isospin $0$ and $1$ channels, respectively,
 and $\delta_{s}=\delta_{0}-\delta_{1}$ is the physical strong phase.
If there is no physical effect of hadronic final state interactions,
 $\delta_{s}=0$ and the triangle of Fig.\ref{fig:TriangleRelation} collapses.
In general
 neglecting final state interactions results
 not only vanishing phases of $\delta_0$ and $\delta_1$,
 but also changing the magnitudes of $| A_0 |$ and $| A_1 |$.
If we truncate the states which contribute to the final state interactions
 by considering only two-body $DK$ states, the relation
\begin{equation}
 \vert \mathcal{M}(\bar{B}^0\to D^+K^-) \vert
 = \vert \mathcal{A}(\bar{B}^0\to D^+K^-) \vert_{\delta_{1,0}=0}
 = \frac{1}{2}(|A_1|+|A_0|)
\label{eq:tree}
\end{equation}
 is satisfied,
 because for each isospin channel there is only one final state.
If we include further the states like $DK \pi \pi$, for example,
 the effect of final state interactions can not be represented only by simple phases
 and the magnitudes of $| A_0 |$ and $| A_1 |$
 are also affected \cite{Suzuki:2007je}.
This truncation of the states,
 or neglecting inelastic final state interactions,
 is the main theoretical assumption in our method, except for isospin symmetry.

There is no justification of this assumption,
 since it has been known that the inelastic final state interactions is important
 in B decays in general \cite{Donoghue:1996hz,Gronau:1998un}.
To be precise we need to describe $\alpha_0|A_0|$ and $\alpha_1|A_1|$
 instead of naive $|A_0|$ and $|A_1|$ in eq.(\ref{eq:tree}),
 where $\alpha_0$ and $\alpha_1$ parametrize the changes of magnitudes of the amplitudes
 by neglecting the effects inelastic final state interactions.
A rough estimate $\alpha_0 \sim \alpha_1 \sim 0.8$ can be obtained
 by using the results of a global fit of the amplitudes and strong phases
 in \cite{Chiang:2007bd}, which means about $20$\% errors in our final results.
This is a large error
 which is comparable to the error from the present measurements of branching fractions.
We certainly need to discover some methods
 to calculate $\alpha_0$ and $\alpha_1$ from the first principle,
 but we leave this task for a future work
 because of the large experimental errors in the measurements of branching fractions
 at this moment in time.
Considering the other way around,
 if the value of $|V_{cb}|$ will be precisely extracted by other methods,
 our method will give a good place to investigate the final state interactions
 in two-body hadronic B decays.

Once the formula of eq.(\ref{eq:tree}) has been accepted,
 we can extract $|V_{cb}V_{us}^*|$ from the values of $|A_0|$ and $|A_1|$
 which, as well as $\cos\delta_s$, can be extracted
 from the measurements of three decay rates.

Now we are going to extract $|V_{cb}|$
 from the experimental values of decay fractions of corresponding three decay modes.
From eqs.(\ref{eq:amplitudes-1}), (\ref{eq:amplitudes-2}) and (\ref{eq:amplitudes-3})
 the ratios of decay fractions can be described as 
\begin{equation}
 \mathcal{R}_1
 \equiv \frac{\mathcal{B}(\bar{B}^0\to D^+K^-)}{\mathcal{B}(B^-\to D^0K^-)}
 = \frac{K_1}{4}
   \left(
    1+ 2 \left|\frac{A_0}{A_1}\right|\cos\delta_s + \left|\frac{A_0}{A_1}\right|^2
   \right),
\label{eq:R1}
\end{equation}
\begin{equation}
 \mathcal{R}_2
 \equiv \frac{\mathcal{B}(\bar{B}^0\to D^0\bar{K}^0)}{\mathcal{B}(B^-\to D^0K^-)}
 = \frac{K_2}{4}
   \left(
    1- 2 \left|\frac{A_0}{A_1}\right|\cos\delta_s + \left|\frac{A_0}{A_1}\right|^2
   \right),
\label{eq:R2}
\end{equation}
 where the coefficients $K_1$ and $K_2$ are kinematical factors of
\begin{equation}
 K_1 = \frac{\tau_{\bar{B}^0} m_{B^-}}{\tau_{B^-}m_{\bar{B}^0}} \cdot
  \frac{
  \sqrt{[1-(m_{D^+}/m_{\bar{B}^0} + m_{K^-}/m_{\bar{B}^0})^2]
        [1-(m_{D^+}/m_{\bar{B}^0} - m_{K^-}/m_{\bar{B}^0})^2]}    
  }
  {
  \sqrt{[1-(m_{D^0}/m_{B^-} + m_{K^-}/m_{B^-})^2]
        [1-(m_{D^0}/m_{B^-} - m_{K^-}/m_{B^-})^2]}
  }, 
\end{equation}
\begin{equation}
 K_2 = \frac{\tau_{\bar{B}^0} m_{B^-}}{\tau_{B^-}m_{\bar{B}^0}} \cdot
  \frac{
  \sqrt{[1-(m_{D^+}/m_{\bar{B}^0} + m_{\bar{K}^0}/m_{\bar{B}^0})^2]
        [1-(m_{D^+}/m_{\bar{B}^0} - m_{\bar{K}^0}/m_{\bar{B}^0})^2]}
  }
  {
  \sqrt{[1-(m_{D^0}/m_{B^-} + m_{K^-}/m_{B^-})^2]
        [1-(m_{D^0}/m_{B^-} - m_{K^-}/m_{B^-})^2]}
  }. 
\end{equation}
Eqs.(\ref{eq:tree}), (\ref{eq:R1}) and (\ref{eq:R2})
 are used to describe $|A_0|$ and $|A_1|$ in terms of $|\mathcal{M}(\bar{B}^0\to D^+K^-)|$ as
\begin{equation}
 |A_0| =  \frac{2}{1+H^{-1}}  |\mathcal{M}(\bar{B}^0\to D^+K^-)|,
\end{equation}
\begin{equation}
 |A_1| =  \frac{2}{1+H} |\mathcal{M}(\bar{B}^0\to D^+K^-)|,
\label{eq:A1}
\end{equation}
 where
\begin{equation}
 H = \left|\frac{A_0}{A_1}\right| = \sqrt{2(\mathcal{R}'_1+\mathcal{R}'_2)-1},
\end{equation}
 and $\mathcal{R}'_i \equiv \mathcal{R}_i/K_i$ with $i=1,2$.
From eqs.(\ref{eq:R1}) and (\ref{eq:R2})
 $\cos\delta_s$ is described only by directly observable quantities as 
\begin{equation}
 \cos\delta_s = \frac{\mathcal{R}'_1-\mathcal{R}'_2}{H}.
\end{equation}
From eq.(\ref{eq:A1})
 the absolute value of the amplitude
 $\vert \mathcal{A}(B^-\to D^0K^-) \vert = |A_1|$ is given by
\begin{equation}
 |\mathcal{A}(B^-\to D^0K^-)|
  = |V_{cb}V_{us}^*| |\mathcal{M}'| \left| \frac{2}{1+H} \right|,
\end{equation}
 where
\begin{equation}
 |\mathcal{M}'|
  = \frac{ |\mathcal{M}(\bar{B}^0\to D^+K^-)| }{ |V_{cb}V_{us}^*| } 
  = \eta_{\text{EW}} \frac{G_{\text{F}}}{\sqrt{2}} a_1(\mu)
    (m_{\bar{B}^0}^2 - m_{D^+}^2) f_{K^{\pm}} f_0(m_{K^{\pm}}^2).
\end{equation}
 is a known quantity.
Finally, we get $|V_{cb}V_{us}^*|^2$
 from the above equation and the value of decay rate $\Gamma(B^-\to D^0K^-)$ as
\begin{equation}\label{eq:V_cb}
|V_{cb}V_{us}^*|^2 = \frac{4\pi m_{B^-}\Gamma(B^-\to D^0K^-)}{\sqrt{ [1-(r_1+r_2)^2][1-(r_1-r_2)^2] }} \frac{|1+H|^2}{|\mathcal{M'}|^2},
\end{equation}
where $r_1=m_{D^0}/m_{B^-}$ and $r_2=m_{K^-}/m_{B^-}$.
This equation is used to extract $|V_{cb}V_{us}^{\ast}|^2$ from experimental data.

For $B \to DK^*$ and $B \to D^*K$, we can extract $|V_{cb}V_{us}^*|^2$ in the same way.
The only major differences are the concrete forms of the amplitudes
 $\mathcal{M}(\bar{B}^0\to D^+K^{\ast-})$ and $\mathcal{M}(\bar{B}^0\to D^{\ast+}K^-)$.

For $\bar{B}^0\to D^+K^{*-}$,
\begin{eqnarray}
 \mathcal{M}(\bar{B}^0\to D^+K^{\ast -})
  &=& \eta_{\text{EW}}\frac{G_\text{F}}{\sqrt{2}} V_{cb}V^*_{us} a_1(\mu)
   \bra{D^{+}(p')}  \bar{c} \gamma_{\mu}(1-\gamma_5)b \ket{\bar{B}^0(p)}
\nonumber \\
  && \quad \times  \bra{K^{\ast-}(p_K)} \bar{s}\gamma^\mu(1-\gamma_5)u \ket{0}
\end{eqnarray}
 with the factorization procedure. 
The first matrix element in the amplitude is given in eq.(\ref{eq:FormFactors}).
The second matrix element in the amplitude is simply described as
\begin{equation}
 \bra{0}\bar{u}\gamma_{\mu}(1-\gamma_5)s \ket{K^*(p_K)} = m_{K^*}  f_{K^{\ast\pm}} \epsilon_\mu (p_{K^*}) ,
\end{equation}
 where $f_{K^{\ast\pm}}$ and $\epsilon_\mu (p_{K^*})$ are the decay constant
 and the polarization vector of $K^{\ast\pm}$ mesons, respectively.
The polarization vector $\epsilon_\mu (p_{K^*})$
 satisfies $\epsilon (p_{K^*})\cdot p_{K^\ast}=0$.
Then, we have
\begin{equation}
 |\mathcal{M}(\bar{B}^0\to D^+K^{\ast-})|
  = \eta_{\text{EW}}\frac{G_{\text{F}}}{\sqrt{2}} |V_{cb}V_{us}^*| a_1(\mu)
    2m_{\bar{B}^0} p^{\ast} f_{K^{\ast\pm}} f_{+}(m_{K^\ast\pm}^2).
\label{eq:amplitudeBDK*}
\end{equation}
Notice that
 this amplitude depends only on the form factor $f_{+}(q^2)$
 instead of $f_{0}(q^2)$ in case of $B \to DK$.
 
For $\bar{B}^0\to D^{\ast+}K^{-}$, the amplitude is given by
\begin{eqnarray}
 \mathcal{M}(\bar{B}^0\to D^{\ast+}K^{-})
  &=& \eta_{\text{EW}}\frac{G_\text{F}}{\sqrt{2}} V_{cb}V^*_{us} a_1(\mu)
  \bra{D^{\ast+}(p')}  \bar{c} \gamma_{\mu}(1-\gamma_5)b \ket{\bar{B}^0(p)}
\nonumber \\
  && \quad \times \bra{K^{-}(p_K)} \bar{s}\gamma^\mu(1-\gamma_5)u \ket{0}
\end{eqnarray}
 with the factorization procedure.
The first matrix element in this amplitude is described as~\cite{Richman:1995wm}
\begin{eqnarray}
 && \bra{D^{\ast +}(p')} \bar{c} \gamma_{\mu}(1-\gamma_5)b\ket{\bar{B}^0(p)} \nonumber \\
 && = \frac{2i\epsilon^{\mu\nu\alpha\beta}}{m_{B}+m_{D^{\ast}}}
       \epsilon^*_{\nu} p'_{\alpha} p_{\beta} V(q^2)
    - (m_{B}+m_{D^{*}}) \left( \epsilon^{\ast\mu}
    - \frac{\epsilon^{*}\cdot q}{q^2} q^{\mu} \right) A_{1}(q^2)
\nonumber \\
 && \quad + \frac{\epsilon^{*}\cdot q}{m_{B}+m_{D^{\ast}}}
    \left[ (p+p')^{\mu} - \frac{m_{B}^2 - m_{D^{\ast}}^2}{q^2} q^{\mu} \right] A_{2}(q^2)
\nonumber \\
 && \quad - 2m_{D^{\ast}} \frac{\epsilon^{*}\cdot q}{q^2} q^{\mu} A_0(q^2),
\end{eqnarray}
 where $\epsilon_\mu (p')$
 is the polarization vector of $D^{\ast}$ meson satisfying $\epsilon(p')\cdot p'=0$
 and $V(q^2)$, $A_{1}(q^2)$, $A_{2}(q^2)$ and $A_{0}(q^2)$ are form factors.
Even though there are many form factors,
 we have a simple expression as
\begin{equation}\label{eq:amplitudeBD*K}
 |\mathcal{M}(\bar{B}^0\to D^{\ast+}K^{-})|
  = \eta_{\text{EW}}\frac{G_{\text{F}}}{\sqrt{2}} |V_{cb}V_{us}^*| a_1(\mu)
    2m_{\bar{B}^0} p^{\ast} f_{K^{\pm}} A_{0}(m_{K^{\pm}}^2).
\end{equation}
Notice that this amplitude depends on only the form factor $A_{0}(q^2)$ .


\section{Numerical analyses and results}
\label{numerical}
In our analysis
 we use the experimental data,
 masses and branching fractions in \cite{Patrignani:2016xqp}
 and the form factors $f_{0,+}(q^2)$ in \cite{Bigi:2016mdz}.
We use the value of the electroweak correction $\eta_{\text{EW}}=1.0066$ in \cite{Beg:1982ex}
 and the short-distance QCD correction $a_{1}(\mu)=1.038$ at leading order with
 $\Lambda^{(5)}_{\overline{\text{MS}}}=225$ MeV and $\mu=4.0$ GeV \cite{Buchalla:1995vs}.
The accuracy of $a_1(\mu)$ is of the order of 1\%.
We do not consider the effect of isospin symmetry breaking
 expecting that the effect is very small within 1\%.

From $B\to DK$ using eq.(\ref{eq:V_cb}) and the experimental data in Table \ref{table:DKinput},
 we obtain $|V_{cb}|=(37\pm 6)\times10^{-3}$ and $\cos\delta_s = 0.60\pm 0.14$.
Notice that the value of $|V_{cb}|$
 is consistent with that determined by both the inclusive and exclusive methods
 with semi-leptonic decays.
\begin{table}[tbp]
\centering
 \begin{tabular}{lcc} \hline \hline
   	 Input & Value & Reference  \\ \hline
         $\tau_{B^0}$ & $(1.520\pm 0.004)\times 10^{-12}$ s &
          \cite{Patrignani:2016xqp} \\
         $\tau_{B^\pm}$ & $(1.638\pm 0.004)\times 10^{-12}$ s &
          \cite{Patrignani:2016xqp} \\
         $\mathcal{B}(\bar{B}^0\to D^+K^-)$ & $(1.86\pm 0.20)\times 10^{-4}$ &
          \cite{Patrignani:2016xqp} \\
         $\mathcal{B}(\bar{B}^0\to D^0\bar{K}^0)$ & $(5.2\pm 0.7)\times 10^{-5}$ &
          \cite{Patrignani:2016xqp} \\ 
         $\mathcal{B}(B^-\to D^0K^-)$ & $(3.74\pm 0.16)\times 10^{-4}$ &
          \cite{Patrignani:2016xqp} \\
         $|V_{us}|$ & $0.2248 \pm 0.0006$ &
          \cite{Patrignani:2016xqp} \\
         $f_{K^\pm}$ & $155.6\pm 0.4$ MeV &
          \cite{Patrignani:2016xqp} \\
         $f_{0}(m_{K^{\pm}}^2)$ & $0.671\pm 0.012$ &
          \cite{Bigi:2016mdz} \\ \hline \hline
  \end{tabular}
\caption{
Inputs for the determination from $B \to DK$.
}
\label{table:DKinput}
\end{table}
The uncertainty of $|V_{cb}|$ is about 30\%
 which is dominated by the experimental errors of the ratios,
 $\mathcal{B}(\bar{B}^0\to D^+K^{-})/\mathcal{B}(B^-\to D^0K^-)$ and
 $\mathcal{B}(\bar{B}^0\to D^{0}\bar{K}^0)/\mathcal{B}(B^-\to D^0K^-)$.
Table \ref{table:DKerror} shows sources of uncertainty of $|V_{cb}|$.
\begin{table}[tbp]
\centering
\begin{tabular}{lc}
 \hline \hline
 Error Source & uncertainty [\%]
 \\
 \hline
 $\Gamma(B^-\to D^0K^-)$ & $4.2$
 \\
 $H$ & $29.8$
 \\ 
 $f_{0}(m_{K^{\pm}}^2)$ & $3.6$
 \\
 \hline
 $|V_{cb}|$ & 30.4
 \\
 \hline \hline
\end{tabular}
\caption{
Sources of uncertainty of $|V_{cb}|$ from $B \to DK$. 
}
\label{table:DKerror}
\end{table}
We find that
 the precise measurements of branching fractions,
 $\mathcal{B}(\bar{B}^0 \to D^+K^{-})$, $\mathcal{B}(\bar{B}^0 \to D^{0}\bar{K}^0)$
 and $\mathcal{B}(\bar{B}^- \to D^{0}\bar{K}^-)$,
 play an important role in the precise determination of $|V_{cb}|$ in our method.
We note that the value of $|V_{cb}|$
 is determined by using the form factor
 which does not employ the CLN parameterization but the BGL parameterization.
To compare the strong phase shift $\cos\delta_s$
 with the one in the previous work \cite{Chiang:2007bd},
 we convert $\cos\delta_s$ to their $\cos\delta_c$,
 where $\delta_c$ is defined as the phase difference between
 $\mathcal{A}(\bar{B}^0\to D^+K^-)$ and $\mathcal{A}(\bar{B}^0\to D^0 \bar{K}^0)$.
Our result $\cos\delta_c=0.43\pm0.16$
 is consistent with that in \cite{Chiang:2007bd} within errors.
 
From $B\to DK^{*}$
 we can obtain the value of $|V_{cb}|$ and the strong phase $\cos\delta_{s}$ in the same way. 
Using eq.(\ref{eq:amplitudeBDK*})
 and the experimental data in Table \ref{table:DK*input},
 we obtain $|V_{cb}|=(41\pm 7)\times10^{-3}$ and $\cos\delta_s = 0.82\pm 0.20$. 
\begin{table}[tbp]
\centering
 \begin{tabular}{lcc}
  \hline \hline
  Input & Value & Reference
  \\ 
  \hline
  $\mathcal{B}(\bar{B}^0\to D^+K^{\ast-})$ & $(4.5\pm 0.7)\times 10^{-4}$
   & \cite{Patrignani:2016xqp}
  \\
  $\mathcal{B}(\bar{B}^0\to D^0\bar{K}^{\ast0})$ & $(4.5\pm 0.6)\times 10^{-5}$
   & \cite{Patrignani:2016xqp}
  \\ 
  $\mathcal{B}(B^-\to D^0K^{\ast-})$ & $(5.3\pm 0.4)\times 10^{-4}$
   & \cite{Patrignani:2016xqp}
  \\
  $f_{K^{\ast\pm}}$ & $205.6\pm 6.0$ MeV & see text
  \\
  $f_{+}(m_{K^{\ast\pm}}^2)$ & $0.696\pm 0.012$ & \cite{Bigi:2016mdz}
  \\
  \hline \hline
 \end{tabular}
\caption{
Inputs for the determination from $B \to DK^{*}$.
}
\label{table:DK*input}
\end{table}
This value of $|V_{cb}|$ is also consistent with both the inclusive and exclusive results.
Notice that $\cos\delta_s$ is larger ($\delta_s$ is smaller) than that in $B\to DK$.
This suggests that
 the effect of hadronic final state interactions
 between a pseudo-scalar meson and a vector mesons
 is less important than that in case of two pseudo-scalar mesons.
The corresponding value of $\cos\delta_c=-0.07\pm0.28$
 is also consistent with that in \cite{Chiang:2007bd} within errors.
We have used the form factor with the BGL parameterization in \cite{Bigi:2016mdz}. 
The decay constant of charged vector meson $f_{K^{*\pm}}$
 is determined by the branching ratio of $\tau \to K^{\ast-}\nu_{\tau}$ \cite{Becirevic:2003pn}.
Since the branching fraction is described as
\begin{equation}
 \mathcal{B}(\tau \to K^{\ast-}\nu_{\tau})
  = \frac{G_{\text{F}}^2m_{\tau}|V_{us}|^2}{8\pi} \tau_{\tau} m_{K^{\ast\pm}}^2 f_{K^{\ast\pm}}^2
  \left( 1- \frac{m_{\tau}^2}{2m_{K^{\ast}}^2} \right)
  \left( 1+ \frac{m_{K^{\ast\pm}}}{m_{\tau}^2} \right)^2,
\end{equation}
 by using the measured values of
 $\mathcal{B}(\tau \to K^{\ast-}\nu_{\tau})
 =(1.20\pm0.07)\times 10^{-2}$,
 $m_{\tau}=1776.86\pm0.12$MeV and $\tau_{\tau} = (290.3\pm 0.5)\times 10^{-15}$s
 \cite{Patrignani:2016xqp}
 we obtain $f_{K^{\ast\pm}} = 205.6\pm6.0$MeV.

From $B \to D^{\ast}K$, in the same way,
 we obtain $|V_{cb}|= (42\pm 9)\times10^{-3}$ and $\cos\delta_s = 0.80\pm0.19$
 using eq.(\ref{eq:amplitudeBD*K}) and the experimental data in Table \ref{table:D*Kinput}.
\begin{table}[tbp]
\centering
 \begin{tabular}{lcc}
 \hline \hline
  Input & Value & Reference
 \\ \hline
 $\mathcal{B}(\bar{B}^0\to D^{\ast+}K^-)$ & $(2.12\pm 0.15)\times 10^{-4}$
  & \cite{Patrignani:2016xqp}
 \\
 $\mathcal{B}(\bar{B}^0\to D^{\ast0}\bar{K}^{0})$ & $(3.6\pm 1.2)\times 10^{-5}$
  & \cite{Patrignani:2016xqp}
 \\ 
 $\mathcal{B}(B^-\to D^{\ast0}K^{-})$ & $(4.20\pm 0.34)\times 10^{-4}$
  & \cite{Patrignani:2016xqp}
 \\
 $A_{0}(m_{K^{\pm}}^2)$ & $0.622\pm 0.062$ & see text
 \\
 \hline \hline
 \end{tabular}
\caption{
Inputs for determination from $B \to D^{*}K$.
}
\label{table:D*Kinput}
\end{table}
This value of $|V_{cb}|$
 is again consistent with those obtained by inclusive and exclusive determinations within errors. 
The value of strong phase supports the previous suggestion
 that the effect of hadronic final state interactions
 is less important in case with a vector meson in final state.
The corresponding value $\cos\delta_c=0.63\pm0.24$
 is also consistent with that in \cite{Chiang:2007bd} within errors. 
The form factor $A_0(q^2)$ is not given by the BGL parameterization,
 because there are no experimental data of the differential decay rate of
 $B \to D^\ast\tau\nu_{\tau}$
 and also no lattice QCD calculations for the form factor.
We have to use the form factor $A_0$
 which is given by the CLN parameterization instead of the BGL parameterization
 by fully utilizing heavy quark symmetry.
The CLN parameterization based on the heavy quark effective theory gives
\begin{equation}
 A_0(q^2) = \frac{R_{0}(w)}{R_{D^{*}}} h_{A_1}(w),
\end{equation}
 where $R_{D^*} = 2\sqrt{m_{B}m_{D^*}}/(m_{B}+m_{D^{*}})$,
\begin{eqnarray}
 h_{A_1}(w)
  &=& h_{A_1}(1) [1-8\rho_{D^*}^2z + (53\rho_{D^*}^2 - 15)z^2 - (231\rho_{D^*}^2 - 91)z^3 ],
\\
 R_0(w)
  &=& R_{0}(1) - 0.11(w-1) + 0.01(w-1)^2,
\end{eqnarray}
 and $w$ and $z$ are kinetic variables defined as
\begin{eqnarray}
 w &=& \frac{m_{B}^2+m_{D^{*}}^2-q^2}{2m_{B}m_{D^{*}}},
\\ 
 z &=& \frac{\sqrt{1+w}-\sqrt{2}}{\sqrt{1+w}+\sqrt{2}}. 
\end{eqnarray}
The value of $h_{A_1}(1)$ has been obtained by the unquenched lattice QCD calculation \cite{Bailey:2014tva}.
The value of $R_0(1)$ can be obtained
 by using the relation based on heavy quark symmetry \cite{Falk:1992wt,Neubert:1992hb}
\begin{equation}
 R_3(1) \equiv \frac{R_2(1)(1-r)+r[R_0(1)(1+r)-2]}{(1-r)^2} = 0.97,
\end{equation}
 if we know the value of $R_2(1)$, where $r=m_{D^*}/m_{B}$.
The values of $R_2(1)$ and $\rho_{D^*}^2$ are determined
 by Belle collaboration \cite{Abdesselam:2017kjf}
 from semi-leptonic ${\bar B}^0 \to D^{*+} l^- {\bar \nu}_l$ decay as
 $R_2(1) = 0.91 \pm 0.08$ and $\rho^2_{D^{\ast}} = 1.17 \pm 0.15$.
In this way
 we obtain the value $R_{0}(1) = 1.08$ with the uncertainty of $10\%$
 considering unknown $\mathcal{O}(1/m_{c}^2)$ corrections. 
Our results are summarized in Table \ref{table:DKresult}.
\begin{table}[tbp]
\centering
 \begin{tabular}{lcc}
 \hline \hline
 Mode & $\cos\delta_{s}$ & $|V_{cb}| \times 10^3$ 
 \\
 \hline
 $B\to DK$  & $0.60\pm 0.14$ & $37\pm 6$ 
 \\
 $B\to DK^*$  & $0.82\pm 0.20$ & $41\pm 7$ 
 \\
 $B\to D^*K$  & $0.80\pm 0.19$ & $42\pm 9$  \\
 \hline \hline
 \end{tabular}
\caption{
Summary of our results. 
}
\label{table:DKresult}
\end{table}


\section{Conclusions}
\label{conclusions}
We have proposed a method of
 extracting the value of $|V_{cb}|$ from hadronic two-body B meson decays.
The recent precise determination
 of the form factor $f_{0}(q^2)$ of semi-leptonic B meson decays in \cite{Bigi:2016mdz}
 allows us to perform this method with $B \to DK$ decay processes.
The main theoretical assumption in our method, except for isospin symmetry,
 is that the effect of {\it inelastic} final state interactions is small.
The small effect of non-factorizable spectator quark scattering has also been neglected,
 which should be included in case with more precise experimental data.
Specifically,
 we have neglected the possible states
 except for $DK$ two-body states in final state interactions.
The quantitative investigation of this truncation is a future work
 which belongs to the efforts to understand non-perturbative QCD physics
 in hadronic decays.
The effect of isospin symmetry breaking is not included,
 since it is negligibly small in the present precision of experimental data.
In future 
 when the errors of branching fractions will be smaller
 and close to 1\% accuracy as well as relevant form factors,
 we need to include the effect of isospin symmetry breaking.
We have used form factors of semi-leptonic B meson decays
 which are determined by using the BGL parameterization in \cite{Bigi:2016mdz,Bigi:2017njr}
 for the extraction of $|V_{cb}|$ from $B \to DK$ and $B\to DK^{\ast}$.
In the extraction of $|V_{cb}|$ from $B \to D^{\ast}K$
 we had to use the CLN parameterization and heavy quark symmetry
 to obtain the form factor $A_0(q^2)$,
 which may contain possibly large uncertainties
 from higher order corrections in heavy quark expansions.

Our final results are summarized in Table \ref{table:DKresult}. 
The extracted values of $|V_{cb}|$ have about $30\%$ uncertainties
 and they are consistent with the values
 from both inclusive and exclusive semi-leptonic decays within errors. 
These consistent results show that our method is reasonable at least in the present precision.
The experimental errors of the hadronic branching fractions,
 in particular $\mathcal{B}(\bar{B}^0\to D^0\bar{K}^0)$,
 $\mathcal{B}(\bar{B}^0\to D^0\bar{K}^{\ast0})$ and
 $\mathcal{B}(\bar{B}^0\to D^{\ast0}\bar{K}^0)$,
 dominate the uncertainty of $|V_{cb}|$. 
We can expect that
 the uncertainty becomes smaller by the results of future experiments and lattice calculations.
It may be possible that
 this method will be the third one competing conventional and established methods
 from inclusive or exclusive semi-leptonic B decays,
 if the problem of {\it inelastic} final state interaction is appropriately treated.

We have also examined
 the effects of hadronic final state interactions in two-body hadronic decays.
The extracted strong phase shifts
 are consistent with the previous works of \cite{Chiang:2007bd,Xing:2003fe,Kim:2004hx}.
The strong phase in $B \to DK$ is larger
 than that in $B\to DK^{\ast}$ and $B\to D^{\ast}K$
 which involve the vector meson in final states (see Table \ref{table:DKresult}).
It is known in general that
 the final state interaction is more important for $B\to PP$ decays
 than $B\to PV$ decays, where $P$ and $V$ indicate pseudo-scalar and vector mesons. 
Here, we must note that the definition of our phases
 are not exactly the same in \cite{Chiang:2007bd,Xing:2003fe,Kim:2004hx},
 and they coincide in the limit of negligible contribution of
 {\it inelastic} final state interactions.
This fact will give a way to investigate
 the magnitude of the effect of {\it inelastic} final state interactions in future.
If the magnitude of $|V_{cb}|$ will be precisely extracted by other methods in future,
 our method will give a good place to investigate the final state interactions
 in two-body hadronic B-decays.


\section*{Acknowledgments}
We would like to thank H.~Kakuno for helpful information about
 SuperKEKB and Belle I\hspace{-.1em}I experiment.
K.M. and Y.S. were supported in part
 by the scholarship of Tokyo Metropolitan University for graduate students.


\end{document}